\documentclass[12pt,preprint]{aastex}
\usepackage{natbib}

\shorttitle{Orbital Motion of Kelu-1AB}
\shortauthors{Gelino, Kulkarni, \& Stephens}

\begin{document}

\title{Evidence of Orbital Motion in the Binary Brown Dwarf Kelu-1AB}
\author{Christopher R. Gelino, S. R. Kulkarni}
\affil{California Institute of Technology, Geological and Planetary Sciences, MC 150-21, 1200 E. California Blvd., Pasadena, CA 91125}
\email{cgelino@gps.caltech.edu, srk@astro.caltech.edu}
\and
\author{Denise C. Stephens}
\affil{Johns Hopkins University, Dept. Physics and Astronomy, Baltimore, MD 21218}
\email{stephens@pha.jhu.edu}


\begin{abstract}
We have resolved Kelu-1 into a binary system with a separation of $\sim$290 mas using the Laser Guide Star Adaptive Optics system on the Keck II telescope.  We have also re-analyzed a 1998 HST observation of Kelu-1 and find that the observed PSF is best fit by a binary object separated by 45 mas.  Observations on multiple epochs confirm the two objects share a common proper motion and clearly demonstrate the first evidence of orbital motion.  Kelu-1B is fainter than Kelu-1A by 0.39$\pm$0.01 magnitudes in the $K'$ filter and 0.50$\pm$0.01 magnitudes in the $H$ filter.  We derive spectral types of L2$\pm$1 and L3.5$\pm$1 for Kelu-1A and B, respectively.  The separation of flux into the two components rectifies Kelu-1's over-luminosity problem that has been known for quite some time.  Given the available data we are able to constrain the inclination of the system to $>$81$\degr$ and the orbital period to $\gtrsim$40 years.
\end{abstract}

\section{Introduction}
L dwarfs are a class of objects with spectral types cooler than the latest M dwarfs \citep{krl+99}.  Their population consists of both substellar (i.e. brown dwarfs) and stellar mass objects, with the brown dwarfs being identified by the presence of Li~\small{I} absorption at 6708~\AA in their spectra.  Kelu-1 was one of the first such brown dwarfs discovered.  It was found at a trigonometric distance of 18.7 pc \citep{dhv+02} as part of a proper motion survey \citep*{rla97}.  Since Kelu-1's discovery, more than 400 L dwarfs have been classified\footnote{See http://dwarfarchives.org for a complete list of all known spectroscopically determined L dwarfs.} in large surveys such as the DEep Near-Infrared Survey (DENIS), the 2-Micron All-Sky Survey (2MASS), and the Sloan Digital Sky Survey (SDSS).  A spectral class sequence was established \citep{krl+99,krl+00,mdb+99}, and Kelu-1 was subsequently assigned a spectral type of L2 in the optical \citep{krl+99} and L3 in the near-IR \citep{klf+04}.  These other discoveries provided a baseline against which to compare Kelu-1's other characteristics.  For example, it was quickly realized that Kelu-1 was much brighter than other L2 dwarfs, causing \citet{mdb+99} to first suggest Kelu-1 a binary.  This super-luminosity has been seen in other studies \citep{lag+01,glm+04} where the main source of this over-brightness was attributed to either an unresolved binary or a young age.

\citet*{mbb99} observed Kelu-1 on 1998 August 14 with the NICMOS camera on HST and no companions were found for separations greater than 300 mas and magnitude differences $<$6.7 mag (F165M filter).  \citet{kkm+99} observed Kelu-1 in a non-AO, seeing-limited program on Keck I using the NIRC instrument on 1998 February 14 and 1999 February 9.  Because Kelu-1 had a fairly circular point spread function during these observations (PSF FWHM$\approx$0.27\arcsec), it was used as a PSF star for the other companion search targets of their project.  While these observations did not discount Kelu-1 as being a very close, unresolved binary, they did warrant a look at the young age hypothesis.  

Based on the strength of the Li I absorption feature, \citet{bmr+98} constrained the age of Kelu-I to be 0.3-1 Gyr.  With this age range and Kelu-1's brightness, it should have an effective temperature of 2100--2350 K, about 400 K higher than objects with similar spectral types \citep{glm+04}.  For it to have an effective temperature consistent with other L3 dwarfs, \citeauthor{glm+04} indicate Kelu-1 must have an age of $\sim$10 Myr, more than an order of magnitude lower than the Li~\small{I}-established lower age limit.  These discrepancies mean that a young age is unlikely to be the cause of the over-luminosity.

Aside from the luminosity, there is other evidence that supports the binarity of Kelu-1.  It is reported to be photometrically and spectroscopically variable with a period of 1.8 hrs \citep*{ctc02,cth03}.  If this period corresponds to an orbital period, then the system would likely be circularized and the orbital radius of the system would be less than a solar radius \citep{ctc02}.  In addition, \citet{ctc02} claim that the $\sim$120 km s$^{-1}$ radial velocity produced by a very close binary such as this could account for the large line broadening (60$\pm$5 km s$^{-1}$) measured by \citet{bha+00}.

We are conducting an adaptive optics (AO) survey of L and T dwarfs at Keck Observatory in an effort to determine binarity in these objects (an article describing the full details of this project is in preparation).  Because of the intrinsic faintness of L and T dwarfs \citep[magnitudes for Kelu-1 are $V$=22.1 and $R$=19.2;][]{rla97}, it is not possible to use them as Natural Guide Stars (NGS) where the magnitude limit is about $V$=14.  Instead, we utilize the newly commissioned Laser Guide Star Adaptive Optics system (LGS AO)\footnote{http://www2.keck.hawaii.edu/optics/lgsao} where the wavefront corrections are performed on a layer of sodium atoms 80 km high in the Earth's mesosphere that are excited by a 10-14 Watt laser tuned to 598nm.  The projected laser and returned light propagate along the same path, requiring the use of a natural star (or other point-like astronomical source) for tip-tilt measurements.  The tip-tilt reference can be as faint as $R$=18.5, but must be within 60\arcsec\ of the science target.  Thus, the area of sky reachable with AO is greatly increased by the less stringent requirements for a reference star provided by the used of the LGS, and many L and T dwarfs binaries can be observed with high resolution imaging from the ground.  We included Kelu-1 in our survey in an effort to resolve the controversy over its binarity and over-luminosity and to search for any low-mass companions.  In this article we discuss the detection of a companion to Kelu-1, first reported in \citet{ll05}, and show the first evidence of orbital motion of the companion.

\section{Observations and Data Reduction}
\subsection{Keck LGS AO}
Observations of Kelu-1 were obtained with NIRC2 (K. Matthews et al. in preparation) behind the LGS AO system on 2005 March 4 and 2005 April 30 during shared-risk time.  A suitable tip-tilt reference star (ID=0643-0289866, $R$=14.3, separation=27\arcsec) was identified in the USNO-B 1.0 catalog \citep{mlc+03}.  Skies were photometric on both nights and the AO system provided very good corrections on our targets.  Kelu-1 was observed with the $K'$ filter during both epochs, with additional exposures in the $H$ filter during follow-up observations in April.  These filters were chosen to maximize the Strehl ratio in our images.  The data were reduced using custom and public IDL scripts.

For each night, two sets of reduced images were produced from the data: a photometric set and an astrometric set.  The only difference in these sets is that an image distortion correction was performed on the astrometric data set.  The distortion in the narrow camera is negligible within a region a few hundred pixels on a side around the center of the chip and gradually increases to as much as 4-5 pixels near the edge.  While the correction fixes the distortion with residuals of 0.81 and 0.62 pixels in the x and y directions, respectively, it does not preserve flux information.  Consequently, its use in images for photometric analysis is not suitable.

The plate scale for both nights was determined using astrometric images of the core of the globular cluster M5.  This field was observed with WFPC2 on HST, with the stars in common with our observations present in the Planetary Camera.  We imaged this field with a 4 (in April) or 5 (in March) position dither pattern in the narrow camera ($\approx$10$\arcsec$ field-of-view).  From the WCS information in the HST image headers, we determined the sky positions of several stars present in our NIRC2 images.  These positions were then used to calculate the plate scale and rotation angle for each position in the dither.  The average of these values is the resultant scale and rotation for that night.  The errors in these quantities are equal to the standard deviation of the measurements.  Table~\ref{scale} presents the plate scale for these nights.  

In our March 4 images of the Kelu-1 field, we saw what appeared to be a binary object with a separation $\sim$300 mas and a differential $K'$ magnitude of $\sim$0.4.  In order to obtain proper motion confirmation of the companion, we re-observed this system on April 30.  On both epochs, we determined the separation and position angle of the candidate companion by using an average of measurements on the individual reduced astrometric images.  The errors are the standard deviations of these measurements.  Table~\ref{astrom} shows these results.  The moderately large proper motion of Kelu-1 \citetext{285.0$\pm$1.0 mas yr$^{-1}$ at 272.2$\pm$0.2$\degr$; \citealp{dhv+02}} made it possible to confidently confirm physical companionship based on observations separated by only two months.  The measured positions of the companion were about 15$\sigma$ away from its predicted background position, clearly demonstrating that these two objects have a common proper motion and are physically associated.  Our binary conclusion firmly supports the recent announcement of Kelu-1's binarity from Keck LGS AO observations obtained in May 2005 \citep{ll05}.

\begin{figure}
\caption{$K'$ image of Kelu-1AB taken on 2005 March 4.  The image is 1.2\arcsec on a side with North up and East to the left.  Kelu-1B is located 284.3$\pm$0.8 mas at a position angle of 220.9$\pm$0.3 deg from Kelu-1A.\label{nirc2}}
\plotone{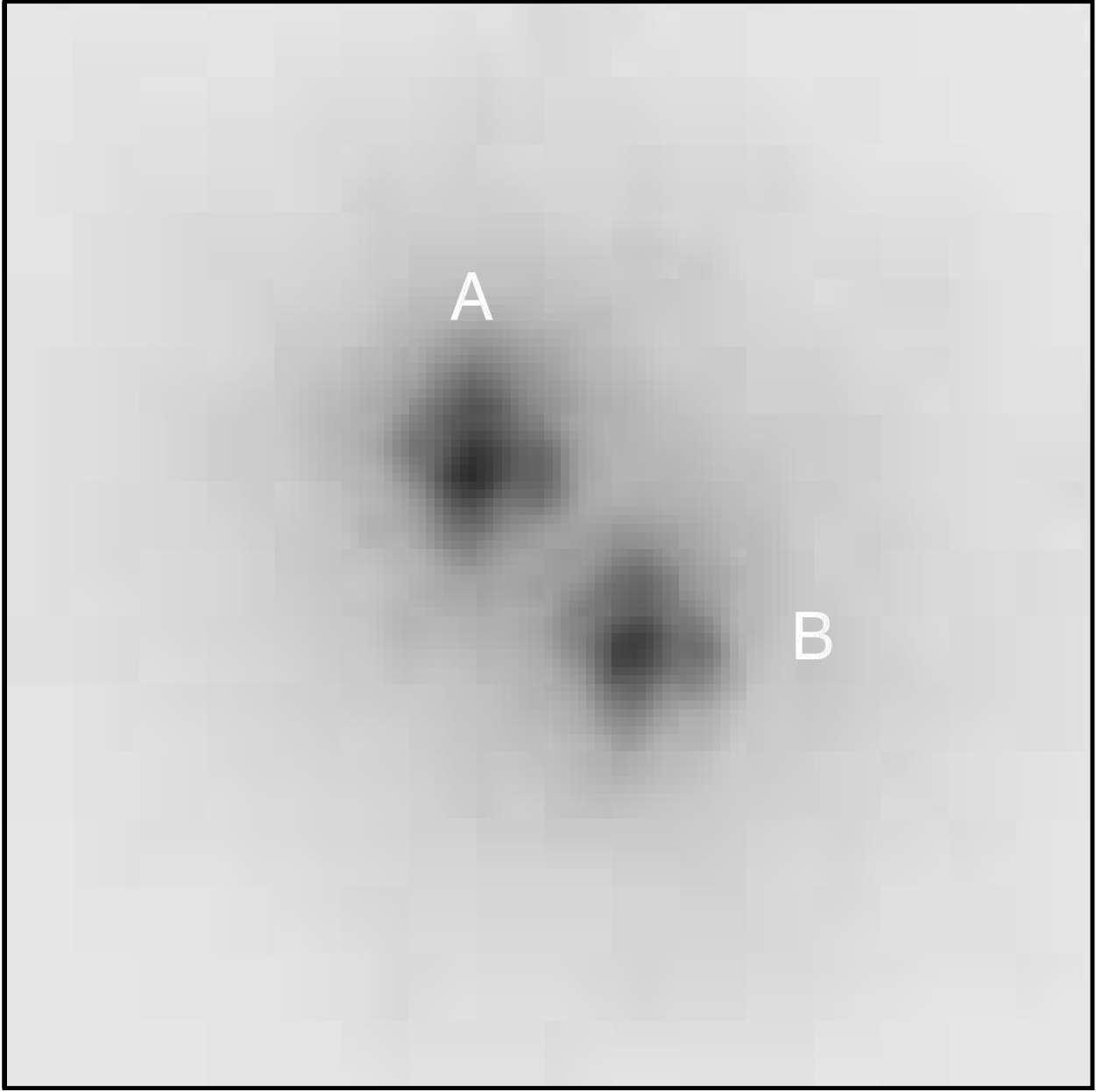} 
\end{figure}

For both nights, all of the photometric images were offset such that northeast object was in the same location in the overlap region.  This image stack was then median averaged to obtain the final image (Figure~\ref{nirc2}), on which the photometry was performed.  The components were well separated and no PSF subtraction routines were needed.  No photometric standards were observed on either night, so only differential photometry is shown in Table~\ref{photom}.  The northeast object is the brighter component in both the $H$ and $K'$.  It will henceforth be called Kelu-1A and the fainter southwest component Kelu-1B.  Our measured astrometry and differential photometry are consistent with those of \citet{ll05}.

\subsection{HST NICMOS}
We re-examined E. Mart{\'i}n's 1998 HST NICMOS images using the PSF modeling technique described in \citet{sn06}.  In short, this technique involves the sub-pixel shifting of two model PSFs and iteratively changing the relative brightness of the PSFs to obtain the minimum residual for each shifted pair.  The lowest minimum residual for all shifted pairs represents the best fit for the observed PSF.  Our analysis discovered that the observed PSFs in all three filters (F110M, F145M, and F165M) were better fit by double PSF solutions than they were by single PSF solutions.  The derived astrometry of the components is presented in Table~\ref{astrom}.  Because the objects are unresolved, a 180$\degr$ ambiguity exists in the value of the position angle.

It is also possible to estimate the magnitudes of the two components (listed in Table~\ref{proptable}) using this technique.  Because the two components are unresolved, the determination of the individual magnitudes is subject to systematic uncertainties in the fitting process.  Therefore, it is difficult to quantify the uncertainty in these magnitudes.  Consequently, although these calculations may not determine the actual difference in magnitudes between the two components, they do indicate the reliability of a binary solution.  Furthermore, the computed magnitudes are also correlated, so if one object is actually fainter than computed, the other component would be brighter, thus conserving the total flux in the system.

\section{Discussion}
\subsection{Component Characteristics}
The spectral types of the components can be estimated using the photometry and absolute magnitude-spectral type relation in \citet{vhl+04} and the differential photometry presented here.  For this exercise we assume that the $\Delta$$H$ and $\Delta$$K'$ measurements in this study are equivalent to $\Delta$$H$ and $\Delta$$K$ in the CIT photometric {\em system} as employed by \citeauthor{vhl+04}  While we know that this is not the case, our interest is only in {\em approximately} determining the spectral types of the components; the {\em only} way to determine spectral types for these or any other objects is to obtain properly calibrated spectra.  Based on the multi-system, near-IR photometry study of \citet{sl04}, we conservatively impose an additional photometric error of 0.15 mag for using this assumption.  This error is added in quadrature to our measured photometric errors.

The $H_{\rm CIT}$ and $K_{\rm CIT}$ apparent magnitudes of Kelu-1AB are 12.46$\pm$0.04 and 11.80$\pm$0.03, respectively \citep{dhv+02}.  Coupled with our measured $\Delta$$H$=0.50$\pm$0.15 and $\Delta$$K'$=0.39$\pm$0.15, the apparent magnitudes of the resolved components become $H$[A]=12.99, $K$[A]=12.37, $H$[B]=13.49, and $K$[B]=12.76 (Table~\ref{proptable}).  Converting these magnitudes from apparent to absolute and applying equations (2) and (3) from \citet{vhl+04} yield a spectral type of $\sim$L2$\pm$1 for Kelu-1A and $\sim$L3.5$\pm$1 for Kelu-1B.  These estimates are consistent with the optical and near-IR spectral types of the composite system and with the spectral types derived by \citet{ll05}.

With this photometry and spectral type estimates, we can place these objects on magnitude-spectral type diagram.  Figure~\ref{magvsst} presents a modified version of Figure~4 from \citet{vhl+04}. In agreement with \citet{ll05}, it is quite clear that the over-luminosity of Kelu-1AB was caused by the unresolved binary and not due to an unusually young effective temperature-based age \citep{glm+04}. 

Finally, we can estimate the effective temperature from the photometry presented here.   \citet{glm+04} calculate a $K$-band bolometric correction (BC$_K$) for Kelu-1AB of 3.31$\pm$0.07.  While this calculation is independent of the spectral type of Kelu-1AB, the spectral type used for Kelu-1AB and all of the other objects in that study were based on near-IR spectra.  The spectral types for Kelu-1A and B presented here are based on a relation derived from optical spectral types.  The near-IR spectral type of Kelu-1AB is L3$\pm$1 \citep{klf+04}, whereas the optical spectral type is L2$\pm$0.5 \citep{krl+99}.  As discussed in \citet{k05}, it is neither uncommon nor unexpected for L and T dwarfs to be classified differently in the optical and near-IR.  Consequently, it is unclear what value of BC$_K$ should be used from \citet{glm+04}.  

Fortunately, for early L dwarfs such as Kelu-1 A and B, the range of BC$_K$ is quite small.  The average value of BC$_K$ for objects with near-IR spectral types L0--L5, inclusive, is 3.32 with a standard deviation of 0.07.  Since this average value is consistent with the value of BC$_K$ computed for the unresolved Kelu-1AB, we choose the unresolved value as the bolometric correction for the individual components.  We note, however, that the uncertainty in the effective temperature determination is largely dominated by the unknown age of Kelu-1 and, given the constraints on the bolometric corrections of early L dwarfs, the exact value of BC$_K$ is unimportant.

Based on our photometry, the absolute $K$ magnitudes of Kelu-1A and B are 11.02 and 11.41, respectively.  Assuming a $K$-band bolometric correction of 3.31 for each component yields bolometric magnitudes of 14.33 and 14.72 and luminosities (log$_{10}$ $L/L_\odot$) of -3.83 and -3.99.  Finally, using an age range of 0.3-1 Gyr \citep{bmr+98} and the \citet{cbah00} DUSTY atmosphere models, we determine an effective temperature range for Kelu-1A of 1900--2100 K and for Kelu-1B 1700--1900 K (Table~\ref{proptable}).  \citet{ll05} computed effective temperatures for Kelu-1A and B (2020 K and 1840 K) by scaling the effective temperature of Kelu-1AB \citep{glm+04} by their luminosity calculations.  Despite the differences between our method and theirs, our results are completely consistent.

\subsection{System Characteristics}
\begin{figure}
\caption{a. Absolute $H$ magnitude as a function of age for Kelu-1A, Kelu-1B (boxed regions) and the DUSTY models (dotted lines) for masses from 0.1 M$_\odot$ (top) to 0.02 M$_\odot$.  b. Same as a except for absolute $K$ magnitude.  The age range is derived from the strength of the Li {\small I} absorption feature in the composite spectrum \citet{bmr+98}.  The combined estimate for the masses of Kelu-1 A and B are 0.060$\pm$0.01 M$_\odot$ and 0.055$\pm$0.01 M$_\odot$, respectively. \label{dusty}}
\plotone{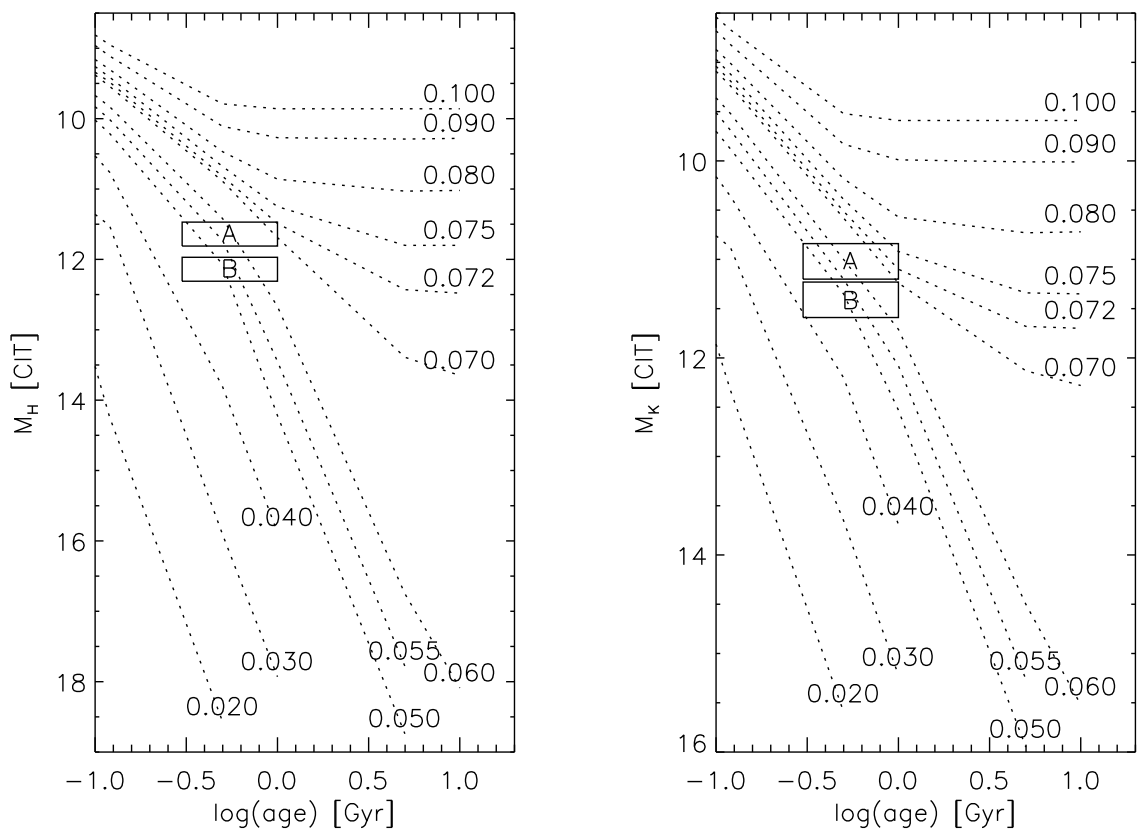}
\end{figure}

\begin{figure} 
\caption{Absolute $K$ magnitude as a function of spectral type for the objects in Table 7 of \citet{vhl+04}.  The position of the unresolved Kelu-1AB is marked as an open square; the individual components are shown with solid squares.  The binarity rectifies the well-documented over-luminosity in the $K$ band.\label{magvsst}}
\plotone{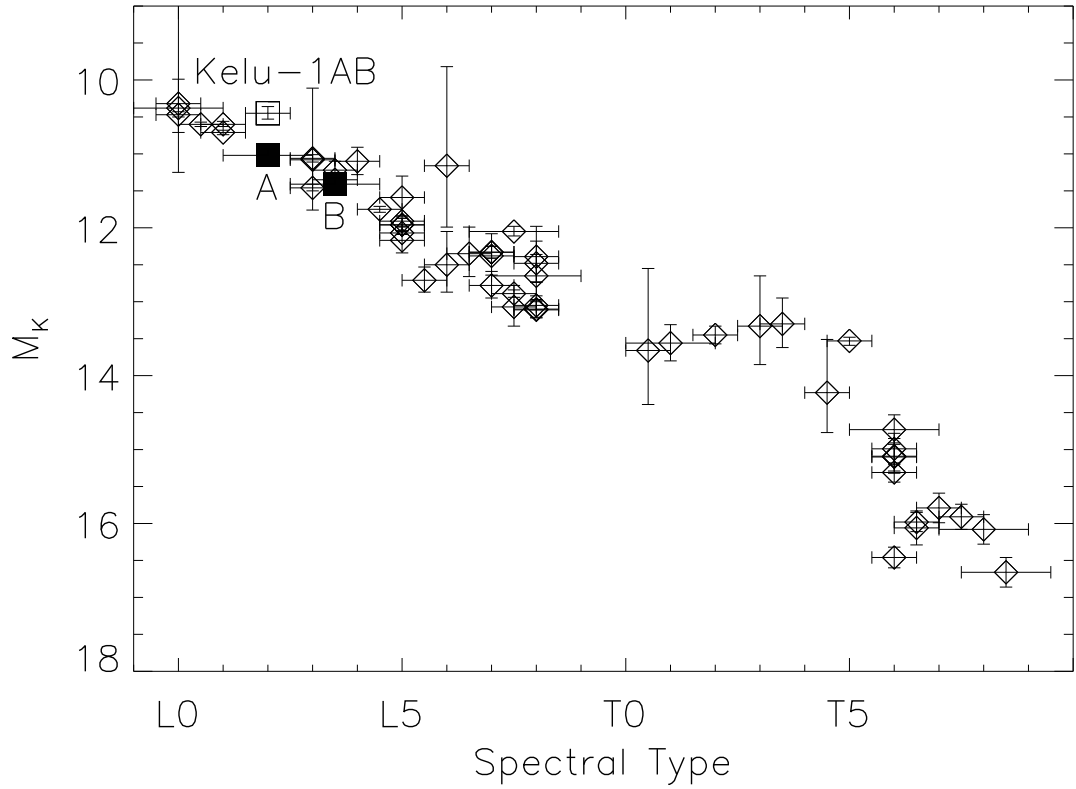}
\end{figure}

It is not known if the Li {\small I} absorption at 6708~\AA~ detected by \citet{rla97} and others is from one or both components.  \citet{ll05} suggest that the presence of any lithium in the unresolved spectrum indicates both components are substellar and, at the very least, Kelu-1B bears lithium in its atmosphere.  Consequently, the masses of both components must be $\lesssim$0.065 M$_\odot$ \citep{cb97,basri98}.  To estimate component masses, we use the photometry derived here along with the age estimate of \citet{bmr+98} to place the system components on the DUSTY atmosphere model evolutionary tracks \citep{cbah00}.  From the combined mass estimates presented in Figure~\ref{dusty}, we estimate the mass of Kelu-1A to be 0.060$\pm$0.01 M$_\odot$ and that of Kelu-1B to be 0.055$\pm$0.01 M$_\odot$, in good agreement with the mass estimates derived by \citet{ll05}.  Even though the conservative errors on these masses do include stellar masses, as mentioned above the true masses are likely to not be greater than about 0.065 M$_\odot$.  Of course, by following the orbital motion of the companion over time, it will be possible to derive dynamical masses, which will provide much needed constraints on the models.

The computation of these dynamical masses will be some time in coming.  Table~\ref{astrom} presents all of the known astrometry of the Kelu-1AB system as of 2006 January.  It is quite clear from this astrometry that Kelu-1B has exhibited orbital motion from the time it was first resolved (our 2005 March 4 observation) to the time of its last observation (a public archival HST observation on 2005 July 31 obtained by W. Brandner).  In fact, the displayed motion suggests that the companion is still moving toward apoapsis.  During this nearly 5 month period the rate of change in the separation has not changed significantly and there is insufficient phase coverage to determine a reliable orbit.  Nonetheless, we can make some crude estimations of the orbit based on the available data.

Given so few points in the orbit, it is impossible to determine how much of the nearly linear motion is caused by a high inclination, very high eccentricity, or both.  In order to simplify the discussion, we assume that the orbit is circular and thus, is being viewed nearly edge-on.   

The most recently measured 2005 positions and the 1998 PSF-modeled position provide some tight constraints on the inclination of the orbit.  Although the position angle of the 4 resolved measurements is not significantly changing, it is not possible to extend a line connecting all four of these points directly back to Kelu-1A.  Therefore, the inclination of the orbit must be $<$90$\degr$.  We can get a rough estimate of the minimum inclination by assuming the maximum extent perpendicular to the 2005 July position is the separation of the modeled PSFs in 1998, i.e. 45 mas.  This assumption results in an inclination $>$81$\degr$ and is much more constrained than the minimum inclination suggested by \citet{ll05} based on the size of the NICMOS PSF.  Finally, since it is quite clear that Kelu-1B is still moving away from Kelu-1A, this minimum inclination estimate will only increase with time.

The period of the binary can be estimated using the 2005 July 31 separation (298 mas or 5.4 AU at the distance of Kelu-1) and the masses derived above.  With the caveat that the most recent separation is not the maximum orbital separation, the calculated period is $\gtrsim37$ years. This period is important for two reasons.  First, it means that it is possible to obtain dynamical masses for the components in a reasonable time-frame, something that has been done for only a few other binary brown dwarf systems \citep{bmb+04,zlp+04,bdk+04}.  Second, this period means that the 1.8 hour period (or 3.6 hours if the variations are ellipsoidal in nature) reported by \citet{ctc02,cth03} is {\em not} due to this binary.  Perhaps one of the components is itself a very close binary, or perhaps the shorter period is simply the rotation period for one of the components.  In addition, the 60 km s$^{-1}$ rotation velocity measured by \citet{bha+00} on 1997 June 2 is from a composite spectrum.  The maximum orbital velocity for Kelu-1AB is only $\sim$3 km s$^{-1}$, which translates to a displacement of $\pm$0.09\AA~ and $\pm$0.08\AA~ for the Cs {\small I} and Rb {\small I} atomics lines used by \citeauthor{bha+00}  This wavelength shift cannot account for the large equivalent widths of these lines (1.7\AA~ and 2.54\AA~ for Cs {\small I} and Rb {\small I}, respectively), so some other mechanism must be at work.  In time it might be possible to separate the lines from each component using high resolution spectroscopy and resolve some of these issues.

\section{Conclusions}
We have used the LGS AO system on the Keck II telescope to observe the binary brown dwarf Kelu-1AB.  We have also re-examined the 1998 HST observations of Kelu-1 and show that the PSFs in those images are best fit by a binary object solution.  Images from multiple epochs confirm the pair as having a common proper motion and demonstrate that Kelu-1B is still heading towards apoapsis.  We constrain the inclination of a circular orbit to $>$81$\degr$.  While this companion detection does rectify the over-luminosity ``problem,'' it cannot account for the 1.8 hour photometric period or the 60 km s$^{-1}$ rotation velocity.  Periodic monitoring of this system with high resolution imaging and spectroscopy is needed to ensure the maximum extent of the orbit is observed and measured, allowing for a more robust determination of the physical orbit.

\acknowledgements{The authors wish to thank the Keck LGSAO Team for their hard work creating such a superb system.  We also thank S. Wiktorowicz, D. Kirkpatrick, and N. Siegler (the referee) for useful discussions, as well as M. B. Stumpf, W. Brandner, \& Th. Henning for informing us of the 2005 HST observation of Kelu-1 through Protostars and Planets V poster \#8571.   This research has benefited from the M, L, and T dwarf compendium housed at DwarfArchives.org and maintained by Chris Gelino, Davy Kirkpatrick, and Adam Burgasser.  The HST M5 calibration data and Kelu-1 images were obtained from the HST data archive at the Space Telescope Science Institute. STScI is operated by the Association of Universities for Research in Astronomy, Inc. under NASA contract NAS 5-26555.  The principle data presented herein were obtained at the W.M. Keck Observatory, which is operated as a scientific partnership among the California Institute of Technology, the University of California and the National Aeronautics and Space Administration. The Observatory was made possible by the generous financial support of the W.M. Keck Foundation.  The authors wish to recognize and acknowledge the very significant cultural role and reverence that the summit of Mauna Kea has always had within the indigenous Hawaiian community.  We are most fortunate to have the opportunity to conduct observations from this mountain.}

\pagebreak

\begin{deluxetable}{lccc}
\tablecaption{NIRC2 Plate Scales and Rotation Angles\label{scale}}
\tablehead{
	\colhead{Date} & \colhead{Scale}       & \colhead{Rotation} & \colhead{Dither} \\
	\colhead{(UT)} & \colhead{(mas/pixel)} & \colhead{(deg)}    & \colhead{Positions} }
\tablewidth{0pt}
\startdata
2005 Mar 4  & 9.97$\pm$0.02 &  +0.08$\pm$0.04 & 5 \\
2005 Apr 30 & 9.96$\pm$0.02 &  +0.11$\pm$0.03 & 4 \\
\enddata
\end{deluxetable}

\pagebreak

\begin{deluxetable}{lccc}
\tablecaption{Kelu-1B Astrometry\label{astrom}}
\tablecolumns{8}
\tablehead{
	\colhead{Date} & \colhead{Separation} & 
	\colhead{P.A.} & \colhead{Reference} \\
	\colhead{(UT)} & \colhead{(mas)} & 
	\colhead{(deg)} & \colhead{} }
\tablewidth{0pt}
\startdata
1998 Aug 14 & 45$\pm$18     & [38.0 or 218.0]$\pm$11.9\tablenotemark{a}  & this work \\
2005 Mar 4  & 284.3$\pm$0.8 & 220.9$\pm$0.3                              & this work \\
2005 Apr 30 & 289.3$\pm$0.7 & 221.1$\pm$0.1                              & this work \\ 
2005 May 1  & 291$\pm$2     & 221.2$\pm$0.6                              & \citet{ll05} \\
2005 Jul 31 & 298$\pm$3     & 221.3$\pm$0.9                              & this work\tablenotemark{b}
\enddata
\tablenotetext{a}{Because the objects are unresolved, there is a 180$\degr$ ambiguity in the position angle solution.}
\tablenotetext{b}{Based on HST observations by W. Brandner that are present in the public archive}
\end{deluxetable}

\pagebreak

\begin{deluxetable}{lcc}
\tablecaption{Kelu-1AB Photometry\label{photom}}
\tablehead{\colhead{Date (UT)} & \colhead{Filter} & \colhead{$\Delta$mag}}
\tablewidth{0pt}
\startdata
2005 Mar 4  & $K'$ & 0.39$\pm$0.01 \\
2005 Apr 30 & $K'$ & 0.39$\pm$0.01 \\
            & $H$   & 0.50$\pm$0.01 
\enddata
\end{deluxetable}

\begin{deluxetable}{lcc}
\tablecaption{Kelu-1AB Derived Properties\label{proptable}}
\tablecolumns{8}
\tablehead{
	\colhead{Property} & \colhead{A} & \colhead{B} }
\tablewidth{0pt}
\startdata
Spectral Type     & L2$\pm$1       & L3.5$\pm$1     \\
$H$ magnitude     & 12.99$\pm$0.15 & 13.49$\pm$0.15 \\
$K$ magnitude     & 12.37$\pm$0.16 & 12.76$\pm$0.16 \\
F110M magnitude   & 14.30$\pm$0.04 & 15.13$\pm$0.10 \\
F145M magnitude   & 13.83$\pm$0.10 & 14.04$\pm$0.13 \\
F165M magnitude   & 13.03$\pm$0.07 & 13.35$\pm$0.11 \\
Mass (M$_\odot$)  & 0.060$\pm$0.01 & 0.055$\pm$0.01 \\
T$_{eff}$ (K)     & 1900--2100     & 1700--1900     \\
\enddata
\end{deluxetable}

\pagebreak

\end{document}